\documentclass[preprint,preprintnumbers, prd,floatfix, superscriptaddress,nofootinbib] {revtex4-1}

\usepackage{epsfig}
\usepackage{subfigure}
\usepackage{dcolumn}
\usepackage{bm}
\usepackage[usenames ,dvipsnames]{xcolor}
\usepackage{slashed}
\usepackage{graphicx,color}
\usepackage{hyperref}
\newcommand{\vsig}{\mbox{\boldmath$\sigma$\unboldmath}}

\begin{document}
\title{Impact of $N^*$ and $\Lambda^*$ resonances on $CP$ violation
in $\Lambda_b^0$ decays}

\author{Yu-Kuo Hsiao}
\email{Co-first author: yukuohsiao@gmail.com}
\affiliation{School of Physics and Electronic Engineering,
Shanxi Normal University, Taiyuan 030031, China}

\author{Kai-Lei Wang}\email{Co-first author: wangkaileicz@foxmail.com}
\affiliation{Department of Physics,
Changzhi University, Changzhi 046011, China}

\author{Juan Wang}\email{Co-first author: wjuanmm@163.com}
\affiliation{School of Physics and Electronic Engineering,
Shanxi Normal University, Taiyuan 030031, China}
\affiliation{Department of Physics,
Changzhi University, Changzhi 046011, China}

\date{\today}

\begin{abstract}
The four-body decay $\Lambda_b^0\to pK^-\pi^+\pi^-$ has led to
the first observation of baryonic $CP$ violation. However,
the underlying subprocesses $\Lambda_b^0\to N^* M$ and $\Lambda_b^0\to \Lambda^* M$,
as well as the roles of excited nucleon ($N^*$) and hyperon ($\Lambda^*$) resonances,
remain largely unexplored. Within the constituent quark model,
we identify the relevant resonant states contributing to these underlying two-body transitions, 
including $N(1535)$, $N(1520)$, $\Lambda(1670)$, $\Lambda(1690)$,
together with the remaining $1P$-wave baryon states.
We obtain the resonant branching fraction ${\cal B}(\Lambda_b^0\to pK^-\pi^+\pi^-)
=(30.0^{+2.8+4.0}_{-1.3-3.4}\pm1.8)\times10^{-6}$, while the resulting
${\cal A}_{CP}(\Lambda_b^0\to pK^-\pi^+\pi^-)=(3.18\pm0.11\pm0.13\pm0.11)\%$
provides a natural interpretation of the first observed baryonic $CP$ asymmetry.
Our analysis establishes the first comprehensive framework
for quantifying the impact of excited baryon resonances in multi-body beauty-baryon decays,
with the associated mechanism generally applicable to baryonic $CP$ asymmetries.
\end{abstract}

\maketitle
\section{introduction}
$CP$ violation is one of Sakharov’s three essential conditions for explaining
the matter–antimatter asymmetry of the Universe.
Since visible matter consists predominantly of baryons,
its manifestation in the baryon sector is therefore of particular significance.
Beauty-baryon decays ${\bf B}_b\to{\bf B}M$ have been extensively studied
in various theoretical frameworks~\cite{Lu:2009cm,Hsiao:2014mua,He:2015fwa,Zhu:2016bra,
Geng:2021nkl,Hsiao:2017tif,Sinha:2021mmx,Han:2024kgz,
Roy:2020nyx,Dery:2020lbc}, where ${\bf B}$ denotes a ground-state baryon and $M$ a meson,
and the corresponding $CP$ asymmetries (${\cal A}_{CP}$) have been extensively investigated.
However, previous measurements
had not provided conclusive evidence~\cite{LHCb:2024iis,LHCb:2025ozp,pdg}.
This situation changed following the observation of
${\cal A}_{CP}(\Lambda_b^0\to pK^-\pi^+\pi^-)=(2.45\pm0.46\pm0.10)\%$
at the 5.2$\sigma$ significance level~\cite{LHCb:2025ray},
establishing the first discovery of baryonic $CP$ violation.
The corresponding theoretical investigations 
have also attracted considerable attention~\cite{He:2025msg,
Chen:2025puj,Zhang:2025mne,Wang:2024oyi,Zhang:2025jnw,Wang:2024rwf}.

Experimentally, the first baryonic $CP$ asymmetry is in fact extracted from
resonant subprocesses of $\Lambda_b^0\to pK^-\pi^+\pi^-$,
including $\Lambda_b^0\to R(p\pi^+\pi^-)K^-$,
$\Lambda_b^0\to R(p\pi^-)R(K^-\pi^+)$, and $\Lambda_b^0\to R(pK^-)R(\pi^+\pi^-)$,
which can be interpreted in terms of
the underlying two-body transitions $\Lambda_b^0\to N^{*+}K^-$,
$\Lambda_b^0\to N^{*0}\bar K_J^{0}$, and
$\Lambda_b^0\to \Lambda^*M^0_J$, respectively, followed by the strong decays
$N^{*+}\to p\pi^+\pi^-$, $N^{*0}\to p\pi^-$ together with $\bar K_J^0\to K^-\pi^+$, and
$\Lambda^*\to pK^-$ together with $M_J^0\to\pi^+\pi^-$. Here,
$N^*$ and $\Lambda^*$ denote the excited nucleon and hyperon states,
respectively. Experimental kinematic selections further constrain
$M_J^0$ to $\rho^0$, $\omega$, and $f_0/f_0(980)$, while
$\bar K_J^0$ includes $\bar K^{*0}/\bar K^{*0}(892)$ and
$\bar K_0^{*0}/\bar K_0^{*0}(1430)$.

Clearly, the decays $\Lambda_b^0\to N^* M$ and $\Lambda_b^0\to\Lambda^* M$
play a key role in baryonic $CP$ asymmetries.
Nonetheless, they remain largely unexplored~\cite{Geng:2016gul,Hsiao:2016hrk,Shang:2026knt}
due to the limited understanding of the $N^*$ and $\Lambda^*$ resonances
participating in these decays. On the other hand, the constituent quark model (CQM),
through its description of baryon spectroscopy, has led to a much improved understanding
of many excited baryon states~\cite{Zhong:2024mnt,Capstick:2000qj,
Capstick:1986ter,Glozman:1996wq,Melde:2008yr,Ferretti:2011zz,
Koniuk:1979vy,Bijker:2015gyk,Zhao:2007hz,Zhong:2013oqa}.
Consequently, the CQM has been successfully applied
to $\Omega_c^0\to \Omega^{*-}\pi^+$ decays~\cite{Wang:2022zja}, and
studies of $\Omega_{(c)}$ and $\Lambda_c$ spectroscopy
in charmful $\Omega_b$ and $\Lambda_b$
decays, respectively~\cite{Wang:2024ozz,Wang:2026dkd,Wang:2025pcs}.

Motivated by these successful applications, 
we extend the CQM framework to the charmless decays
$\Lambda_b^0\to N^* M$ and $\Lambda_b^0\to \Lambda^* M$.
We seek to establish the first comprehensive theoretical framework 
for systematically identifying the relevant $N^*$ and $\Lambda^*$ resonances, 
thereby enabling a quantitative description of the associated branching fractions 
and baryonic $CP$ asymmetries in multi-body ${\bf B}_b$ weak decays.

\section{Formalism}
%
\begin{figure}[t]
\includegraphics[width=1.4in]{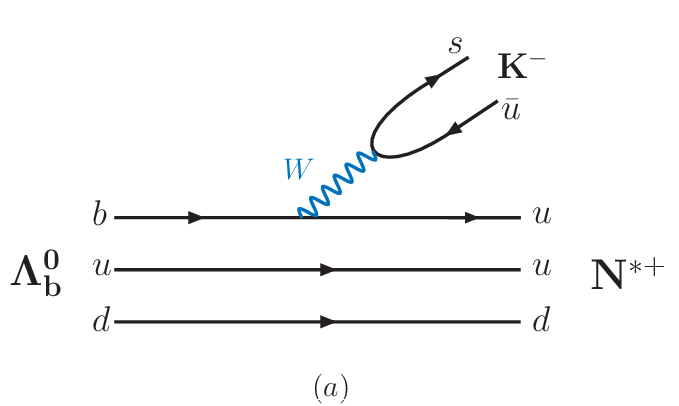}
\includegraphics[width=1.4in]{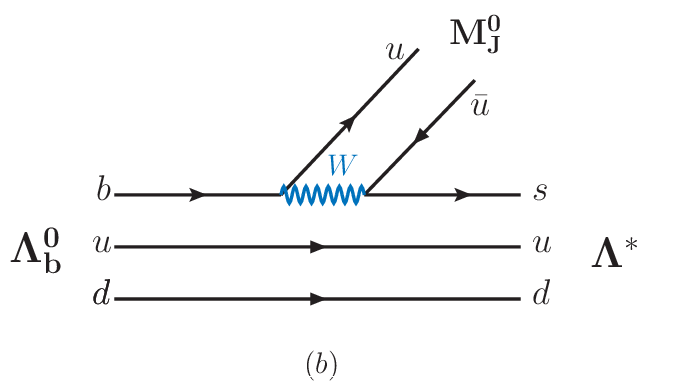}
\includegraphics[width=1.4in]{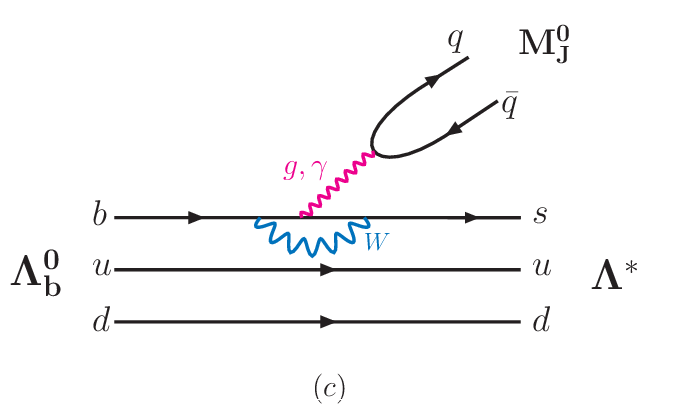}
\includegraphics[width=1.8in]{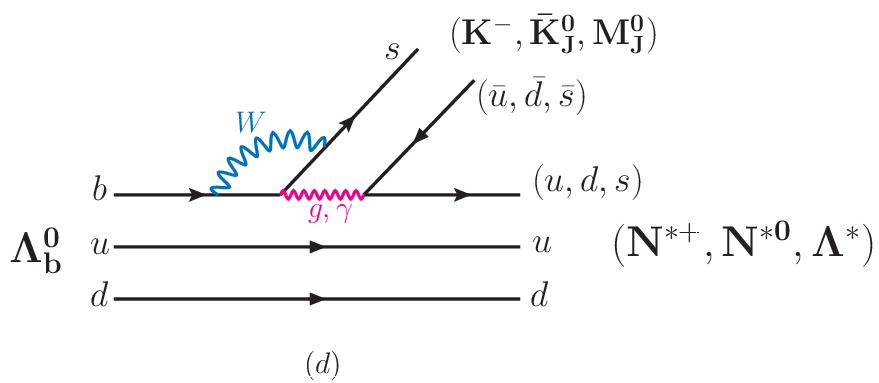}
\caption{Feynman diagrams for $(a, d)$ $\Lambda_b^0\to N^{*+} K^-$,
(d) $\Lambda_b^0\to N^{*0}\bar K_J^0$, and
$(b, c, d)$ $\Lambda_b^0\to \Lambda^* M_J^0$.}\label{fig1}
\end{figure}
%
The two-body decays $\Lambda_b^0\to N^{*+}K^-$, $N^{*0}\bar K_J^0$, and $\Lambda^* M_J^0$,
shown in Figs.~\ref{fig1}(a)--(d), are taken as the underlying subprocesses 
of the resonant decay $\Lambda_b^0\to pK^-\pi^+\pi^-$.
Identifying the relevant $N^*$ and $\Lambda^*$ resonances 
and quantifying their contributions are essential for understanding the first observed baryonic $CP$ asymmetry.
The corresponding amplitudes 
are written as~\cite{Ali:1998eb,Hsiao:2014mua,Hsiao:2017tif,Geng:2016gul,Hsiao:2016hrk}
\begin{eqnarray}\label{amp1}
&&
\hat{\cal M}(\Lambda_b^0\to N^{*+} K^-)=
(\alpha_1^s+\alpha_4^s)\langle K^-|(\bar s u)_{V-A}|0\rangle
\langle N^{*+}|(\bar u b)_{V-A}|\Lambda_b^0\rangle\nonumber\\
&&+\alpha_6^s\langle K^-|(\bar s u)_{S+P}|0\rangle
\langle N^{*+}|(\bar u b)_{S-P}|\Lambda_b^0\rangle\,,\nonumber\\
&&
\hat{\cal M}(\Lambda_b^0\to N^{*0} \bar K_J^0)=
\alpha_4^s\langle \bar K_J^0|(\bar s d)_{V-A}|0\rangle
\langle N^{*0}|(\bar d b)_{V-A}|\Lambda_b^0\rangle
\nonumber\\&&
+\alpha_6^s\langle \bar K_J^0|(\bar s d)_{S+P}|0\rangle
\langle N^{*0}|(\bar d b)_{S-P}|\Lambda_b^0\rangle\,,\nonumber\\
&&
\hat{\cal M}(\Lambda_b^0\to\Lambda^* M_J^0)=
[\alpha_2^s\langle M_J^0|(\bar uu)_{V-A}|0\rangle+
\alpha_3^s\langle M_J^0|(\bar uu+\bar dd+\bar ss)_{V-A}|0\rangle\nonumber\\
&&+
\alpha_4^s\langle M_J^0|(\bar ss)_{V-A}|0\rangle+
\alpha_5^s\langle M_J^0|(\bar uu+\bar dd+\bar ss)_{V+A}|0\rangle+
\alpha_9^s\langle M_J^0|(2\bar uu-\bar dd-\bar ss)_{V-A}|0\rangle]\nonumber\\
&&
\times\langle \Lambda^*|(\bar sb)_{V-A}|\Lambda_b^0\rangle
+\alpha_6^s\langle M_J^0|(\bar ss)_{S+P}|0\rangle\langle
\Lambda^*|(\bar sb)_{S-P}|\Lambda_b^0\rangle\,,
\end{eqnarray}
with ${\cal M}\equiv (G_F/\sqrt 2)\hat{\cal M}$, where $G_F$ is the Fermi constant.
We define $(\bar q_1 q_2)_{V\pm A}\equiv\bar q_1\gamma_\mu(1\pm\gamma_5)q_2$
and $(\bar q_1 q_2)_{S\pm P}\equiv\bar q_1(1\pm\gamma_5)q_2$.
The coefficients $\alpha_i^q$ are given by $\alpha_{1,2}^q=V_{ub}V_{uq}^* a_{1,2}$,
$(\alpha_{3,4,5}^q,\alpha_9^q)=-V_{tb}V_{tq}^*(a_{3,4,5},a_9/2)$,
and $\alpha_6^q=V_{tb}V_{tq}^* 2a_6$, where
$a_i=c_i^{\rm eff}+c_{i\pm1}^{\rm eff}/N_c^{\rm eff}$ for $i=$ odd (even),
$c_i^{\rm eff}$ are the effective Wilson coefficients~\cite{Ali:1998eb,Hsiao:2017tif},
$N_c^{\rm eff}$ is the effective color number, and
$V_{ij}$ are the Cabibbo--Kobayashi--Maskawa (CKM) matrix elements.
Within the generalized factorization framework~\cite{Ali:1998eb}, 
we take $N_c^{\rm eff}=3$ as the central value and vary it over
$2\le N_c^{\rm eff}\le\infty$ to estimate nonfactorizable effects.

The intermediate states $R(p\pi^+\pi^-)$, $R(p\pi^-)$, and $R(pK^-)$
exhibit clear resonant structures in the invariant-mass region $(1.5\text{--}1.8)$~GeV
for the four-body decay $\Lambda_b^0\to pK^-\pi^+\pi^-$. As a consequence,
the $N^*$ and $\Lambda^*$ resonances
involved in the underlying $\Lambda_b^0\to N^* M,\Lambda^* M$ decays
are identified as members of the $1P$-wave nucleon and hyperon states
within the CQM framework,
including $N(1535)/\Lambda(1670)$, $N(1520)/\Lambda(1690)$,
$N(1650)$, $N(1700)$, and $N(1675)$, with quantum numbers
$J^P=(1/2^-,3/2^-,1/2^-,3/2^-,5/2^-)$, 
respectively~\cite{Zhong:2024mnt,Capstick:2000qj,Capstick:1986ter,
Glozman:1996wq,Melde:2008yr,Ferretti:2011zz,Koniuk:1979vy,
Bijker:2015gyk,Zhao:2007hz,Zhong:2013oqa},
together with the $1P$-wave hyperon singlets
$\Lambda(1405)$ and $\Lambda(1520)$, carrying $J^P=(1/2^-,3/2^-)$,
respectively~\cite{Capstick:2000qj,Capstick:1986ter,Melde:2008yr,
Koniuk:1979vy,Bijker:2015gyk,Zhao:2007hz,Zhong:2013oqa}.
A careful identification of the possible contributions from
these $1P$-wave $N^*$ and $\Lambda^*$ resonances therefore
requires the evaluation of a total of 27 decay channels
in $\Lambda_b^0\to N^{*+}K^-$, $N^{*0}\bar K_J^0$, and $\Lambda^* M_J^0$.

\begin{figure}[t]
\includegraphics[width=2.5in]{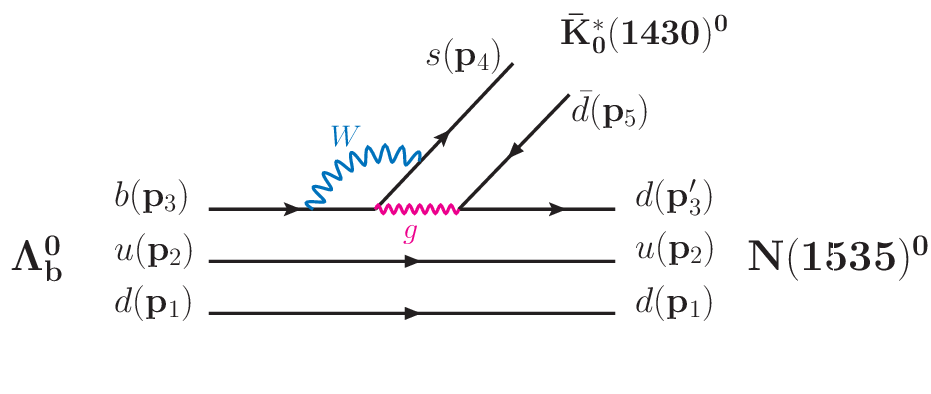}
\caption{The $\Lambda_b^0\to N(1535)^0\bar K_0^*(1430)^0$ process,
with the momenta of the involved quarks explicitly assigned and indicated.}\label{fig2}
\end{figure}

We illustrate the calculation using $\Lambda_b^0\to N_{1535}^0\bar K_0^{*0}$, where
$N_{1535}^0\equiv N(1535)^0$.
The dominant contribution arises from the internal gluon-emission topology shown in Fig.~\ref{fig2}.
Accordingly, the amplitude $\hat{\cal M}(\Lambda_b^0\to N_{1535}^0\bar K_0^{*0})$ in Eq.~(\ref{amp1})
is governed by the effective Hamiltonian for the $b\to sd\bar d$ transition,
${\cal H}_{\rm eff}=\sum_{i=3}^6 c_i O_i$. 
Within the factorization approximation, $O_{3(5)}\simeq O_{4(6)}/N_c^{\rm eff}$,
such that the contributions from $c_{3(5)}O_{3(5)}$ are absorbed into $c_{4(6)}O_{4(6)}$.
The resulting effective operator structure involves only the QCD penguin operators
$O_4=\bar{\psi}_{\bar{s}_\beta}\gamma_\mu(1-\gamma_5)\psi_{d_\beta}
\bar{\psi}_{\bar{d}_\alpha}\gamma^\mu(1-\gamma_5)\psi_{b_\alpha}$ and
$O_6=2\bar{\psi}_{\bar{s}_\beta}(1+\gamma_5)\psi_{d_\beta}
\bar{\psi}_{\bar{d}_\alpha}(1-\gamma_5)\psi_{b_\alpha}$~\cite{Buchalla:1995vs},
where $\psi_{j_\delta}$ represent the $j$-quark field
and $\delta=(\alpha,\beta)$ denotes the color indices.
%
%
In the non-relativistic limit, the operator $O_4$ can be decomposed as $O_4\simeq O_4^{PC}+O_4^{PV}$,
where $O_4^{PC}$ and $O_4^{PV}$ denote the parity-conserving 
and parity-violating components of the weak transition, respectively.
Their explicit forms are~\cite{Wang:2024ozz,Wang:2026dkd,Wang:2022zja}
\begin{eqnarray}\label{O4}
O_4^{PC}&=&
\delta^3(\textbf{p}_3-\textbf{p}_3'-\textbf{p}_4-\textbf{p}_5)/(2\pi)^3\hat{O}_f \hat{O}_c \nonumber\\
&\times\bigg\{& \vsig_4 \cdot \bigg[\left(\frac{\textbf{p}_5}{2m_5}+\frac{\textbf{p}_4}{2m_4}\right)
-\left(\frac{\textbf{p}_3'}{2m_3'}+\frac{\textbf{p}_3}{2m_3}\right)
+i\vsig_3 \times \left(\frac{\textbf{p}_3}{2m_3}-\frac{\textbf{p}_3'}{2m_3'}\right)
\bigg]\nonumber\\
&+&
\vsig_3 \cdot \bigg[\left(\frac{\textbf{p}_3'}{2m_3'}+\frac{\textbf{p}_3}{2m_3}\right)
-\left(\frac{\textbf{p}_5}{2m_5}+\frac{\textbf{p}_4}{2m_4}\right)
+i \vsig_4\times \left(\frac{\textbf{p}_4}{2m_4}-\frac{\textbf{p}_5}{2m_5}\right)
\bigg]\bigg \}\,,\nonumber\\
O_{4}^{PV}&=&
\delta^3(\textbf{p}_3-\textbf{p}_3'-\textbf{p}_4-\textbf{p}_5)/(2\pi)^3
\hat{O}_f \hat{O}_c (\vsig_3 \cdot \vsig_4 - 1)\,,
\end{eqnarray}
where $\mathbf{p}_j$, $m_j$, and $\vsig_j$ 
denote the momentum, mass, and spin of the $j$th quark, respectively, as labeled in Fig.~\ref{fig2}.
The operators $\hat O_f$ and $\hat O_c$ represent the flavor and color structures, respectively.
Specifically, $\hat O_f=b_5^\dagger(s)b_4^\dagger(\bar d)b_3^\dagger(d)b_3(b)$
describes the $b\to d$ transition together with the creation of an $s\bar d$ pair, 
while $\hat{O}_c=\delta_{c_4 c_5}\delta_{c_3' c_3}$ ensures that 
the quark pairs associated with $O_4$ are coupled to color-singlet configurations.
Similarly, $O_6$ can be decomposed as $O_6\simeq O_6^{PC}+O_6^{PV}$.
Inserting $O_{4,6}^{PC(PV)}$ into the amplitude yields
${\cal M}(\Lambda_b^0\to N_{1535}^0 \bar K_0^{*0})=\alpha^s_4({\cal M}^{PC}_4
+{\cal M}^{PV}_4)+\alpha_6^s({\cal M}^{PC}_6+{\cal M}^{PV}_6)$, where
\begin{eqnarray}\label{amp2}
({\cal M}_{4(6)}^{PC,PV})^{J_1^z,J_2^z,J_3^z}_{J_1,J_2,J_3}=
\langle N_{1535}^0(\textbf{P}_2,J_2,J_2^z) \bar K_0^{*0}(\textbf{q},J_3,J_3^z)|
O_{4(6)}^{PC,PV}|\Lambda_b^0(\textbf{P}_1,J_1,J_1^z)\rangle\,.
\end{eqnarray}
In the above equation, $\textbf{q}=\textbf{P}_1-\textbf{P}_2$ is the momentum transfer, and
$\Lambda_b(\mathbf{P}_1,J_1,J_1^z)$, $N_{1535}^0(\mathbf{P}_2,J_2,J_2^z)$,
and $\bar K_0^{*0}(\mathbf{q},J_3,J_3^z)$ denote the wave functions.

The meson wave function is defined as $M(\mathbf{P}_{3},J,J_z)=
\int d  \mathbf{p}_4 d  \mathbf{p}_5 \delta^3( \mathbf{p}_4 + \mathbf{p}_5- \mathbf{P}_3)
\Psi_M(\mathbf{p}_4,\mathbf{p}_5)$ \cite{Wang:2026dkd,Hayne:1981zy}, where
$\Psi_M(\mathbf{p}_4,\mathbf{p}_5)$ denotes the momentum-space
wave function, and $\mathbf{p}_4$ and $\mathbf{p}_5$
are the three-momenta of the constituent quark $q_4$ and antiquark $\bar q_5$,
respectively. In the Jacobi-momentum framework,
the two independent quark momenta are recast into the relative momentum
$\mathbf{p}_M=(\textbf{p}_4-\textbf{p}_5)/\sqrt 2$, such that
the internal motion of the meson is described as a simple harmonic oscillation
of the constituent quark–antiquark system. This leads to the expression
$\Psi_M(\mathbf{p}_4,\mathbf{p}_5)=
\sum_{m_M} \mathcal{C}^{l_M,S,J_3}_{m_M,S_z,J_3^z}
\zeta_{M}\varphi_{M}\chi^S_{S_z}\psi_{n_M l_M m_M}(\mathbf{p}_M)$,
where $\mathcal{C}^{l_M,S,J_3}_{m_M,S_z,J_3^z}
\equiv\langle l_M,m_M;S,S_z|J_3,J_3^z\rangle$ represents
the Clebsch-Gordan coefficient (CGC), and
$\zeta_{M}$, $\varphi_{M}$, and $\chi^S_{S_z}$ denote the color, flavor, spin wave functions,
respectively. Moreover, $\psi_{n_M l_M m_M}(\mathbf{p}_M)$
is presented with the simple harmonic oscillation function,
defined by~\cite{Wang:2025pcs,Wang:2026dkd,Wang:2022zja,Wang:2024ozz}
\begin{eqnarray}\label{SHO}
\psi_{n l m}(\mathbf{p})
&=&(i)^{l}(-1)^{n}
\left[\frac{2n!}{(n+l+1/2)!}\right]^{1/2}\frac{1}{\alpha^{l+3/2}}
\exp\left(-\frac{\mathbf{p}^2}{2\alpha^2}\right)
L_{n}^{l+1/2}\left(\frac{\mathbf{p}^2}{\alpha^2}\right)
\mathcal{Y}_{l m}(\mathbf{p})\,,
\end{eqnarray}
where $n$, $l$, and $m$ correspond to the principal, angular,
and magnetic quantum numbers, respectively,
$L_n^{l+1/2}$ is the associated Laguerre polynomial,
$\mathcal{Y}_{lm}(\textbf{p})=|\textbf{p}|^{l}Y_{lm}(\mathbf{\hat{p}})$
is the solid harmonic constructed from the spherical harmonic $Y_{lm}(\mathbf{\hat{p}})$,
and $\alpha$ the oscillator parameter.

In the case of the scalar meson $\bar K_0^{*0}(s\bar d)$,
the constituent quark-antiquark pair is assigned to the $^3P_0$ configuration,
with $(S,L,J)=(1,1,0)$. The spin-orbital components $(J,J_z)=(0,0)$ are
$(S_z,L_z)=(1,-1)$, $(0,0)$, and $(-1,1)$. The spin wave functions are
$(\chi^1_{1},\chi^1_{0},\chi^1_{-1})
=(\uparrow\uparrow,(\uparrow\downarrow+\downarrow\uparrow)/\sqrt{2},
\downarrow\downarrow)$.
The corresponding orbital components are described by
$\psi_{0\,1\,m_M}(\mathbf{p}_M)$ with $m_M=L_z=(-1,0,1)$.
Using $\mathcal{C}^{1,1,0}_{-1,1,0}=-\mathcal{C}^{1,1,0}_{0,0,0}
=\mathcal{C}^{1,1,0}_{1,-1,0}=1/\sqrt{3}$, together with
$\zeta_{\bar K_0^{*0}}=(R\bar R+G\bar G+B\bar B)/\sqrt{3}$ and
$\varphi_{\bar K_0^{*0}}=s\bar d$, 
we obtain $\Psi_{\bar K_0^{*0}}(\mathbf{p}_4,\mathbf{p}_5)
=\zeta_{\bar K_0^{*0}}\varphi_{\bar K_0^{*0}}
[\chi^1_{-1}\psi_{0\,1\,1}(\mathbf{p}_{\bar K_0^{*0}})
-\chi^1_{0}\psi_{0\,1\,0}(\mathbf{p}_{\bar K_0^{*0}})$
$+\chi^1_{1}\psi_{0\,1\,-1}(\mathbf{p}_{\bar K_0^{*0}})]/\sqrt{3}$~\cite{Pang:2017dlw,
Ebert:2009ub}.

The baryon wave function is defined as
${\bf B}( \mathbf{P}_{\bf B},J,J_z)=
\int d \mathbf{p}_1 d \mathbf{p}_2 d \mathbf{p}_3\,
\delta^3( \mathbf{p}_1 + \mathbf{p}_2+ \mathbf{p}_3- \mathbf{P}_{\bf B})
\Psi_{\bf B}(\mathbf{p}_1,\mathbf{p}_2,\mathbf{p}_3)$~\cite{Wang:2026dkd,Hayne:1981zy},
where $\Psi_{\bf B}(\mathbf{p}_1,\mathbf{p}_2,\mathbf{p}_3)$ denotes the
momentum-space baryon wave function, and
$\mathbf{p}_{1,2,3}$ are the three-momenta of the constituent quarks.
In the Jacobi-momentum framework, these momenta are recombined as
$\mathbf{p}_\rho=(\textbf{p}_1-\textbf{p}_2)/\sqrt 2$ and
$\mathbf{p}_\lambda=\sqrt{3/2}
[m_3(\textbf{p}_1+\textbf{p}_2)-(m_1+m_2)\textbf{p}_3]/(m_1+m_2+m_3)$.
Here, $\mathbf{p}_\rho$ describes the relative motion between $q_1$ and $q_2$,
while $\mathbf{p}_\lambda$ characterizes the motion between $q_3$ and the
$q_1q_2$ subsystem. Accordingly, the baryon is described by the
$\rho$- and $\lambda$-mode oscillators, and
$\Psi_{\bf B}$ contains the two SHO wave functions
$\psi^\rho_{n_\rho l_\rho m_\rho}(\mathbf{p}_\rho)$ and
$\psi^\lambda_{n_\lambda l_\lambda m_\lambda}(\mathbf{p}_\lambda)$.

The $\Lambda_b^0$ baryon is assigned to the lowest $\rho$-$\lambda$ oscillator state,
with $n_i=l_i=m_i=0$ for $i=(\rho,\lambda)$. The $ud$ pair is taken as a spin-zero
diquark, $s_\rho=0$, which couples with the $b$-quark spin to give
$S=1/2$. The corresponding spin wave functions are $\chi^{S=1/2}_{1/2,s_\rho=0}=
(\uparrow\downarrow\uparrow-\downarrow\uparrow\uparrow)/\sqrt{2}$ and
$\chi^{S=1/2}_{-1/2,s_\rho=0}=
(\uparrow\downarrow\downarrow-\downarrow\uparrow\downarrow)/\sqrt{2}$.
Thus, $\Psi_{\Lambda_b^0}(\textbf{p}_\rho,\textbf{p}_\lambda)
= \zeta_{\Lambda_b^0}\phi_{\Lambda_b^0} \chi_{S_z=\pm 1/2,s_\rho=0}^{S=1/2}
\psi^\rho_{000}(\textbf{p}_\rho)\psi^\lambda_{000}(\textbf{p}_\lambda)$~\cite{Wang:2025pcs}, 
where $\zeta_{\Lambda_b^0} =(RGB-RBG+GBR-GRB+BRG-BGR)/\sqrt{6}$
and $\phi_{\Lambda_b^0} =(udb-dub)/\sqrt{2}$.

The $N^0_{1535}$ baryon is assigned as a $1P$-wave excited nucleon
with $J^P=1/2^-$. The orbital excitation with $L=1$ may arise
either from the $\rho$-mode or the $\lambda$-mode oscillator,
giving the possible spatial configurations
$\psi_{01M_L}^{\rho}(\mathbf{p}_\rho)\psi_{000}^{\lambda}(\mathbf{p}_\lambda)$ and
$\psi_{000}^{\rho}(\mathbf{p}_\rho)\psi_{01M_L}^{\lambda}(\mathbf{p}_\lambda)$,
with $M_L=(-1,0,+1)$. For the spin-1 $ud$ subsystem, $s_\rho=1$,
coupling with the third-quark spin gives the total spin $S=1/2$.
The corresponding spin wave-function components are
$\chi^{S=1/2}_{1/2,s_\rho=1}=
-(\uparrow\downarrow\uparrow+\downarrow\uparrow\uparrow-2\uparrow\uparrow\downarrow)
/\sqrt{6}$, and $\chi^{S=1/2}_{-1/2,s_\rho=1}=(\uparrow\downarrow\downarrow+\downarrow\uparrow\downarrow-2\downarrow\downarrow\uparrow)/\sqrt{6}$.
Thus, we write $\Psi_{N_{1535}^0}(\mathbf{p}_\rho,\mathbf{p}_\lambda)
=\sum_{M_L,S_z} \mathcal{C}^{L=1,S=1/2,J=1/2}_{M_L,S_z,J_z} \zeta_{N_{1535}^0} 
[(\phi^\rho \chi^{1/2}_{S_z,1}+\phi^\lambda \chi^{1/2}_{S_z,0})
\psi_{01M_L}^{\rho}(\textbf{p}_\rho)\psi_{000}^{\lambda}(\textbf{p}_\lambda)$
$+(\phi^\rho \chi^{1/2}_{S_z,0}-\phi^\lambda \chi^{1/2}_{S_z,1})
\psi_{000}^{\rho}(\textbf{p}_\rho)\psi_{01M_L}^{\lambda}(\textbf{p}_\lambda)]/\sqrt{2}$~\cite{Zhong:2024mnt,Koniuk:1979vy},
where $\phi^{\rho}=1/\sqrt{2}(udd-dud)$ and $\phi^{\lambda}=1/\sqrt{6}(dud+udd-2ddu)$
denote the mixed-antisymmetric and mixed-symmetric flavor wave functions,
and $\zeta_{N^0_{1535}}=\zeta_{\Lambda_b^0}$.
The relevant CGCs satisfy $M_L+S_z=J_z$ and are
$\mathcal{C}^{1,1/2,1/2}_{0,-1/2,-1/2}=-\mathcal{C}^{1,1/2,1/2}_{0,1/2,1/2}=\sqrt{1/3}$ and
$\mathcal{C}^{1,1/2,1/2}_{-1,1/2,-1/2}=-\mathcal{C}^{1,1/2,1/2}_{1,-1/2,1/2}=-\sqrt{2/3}$.

Substituting the meson and baryon wave functions into Eq.~(\ref{amp2}),
we obtain the nonvanishing amplitudes:
\begin{eqnarray}
(\mathcal{M}_{4,6}^{PC})_{\frac{1}{2},\frac{1}{2},0}^{-\frac{1}{2},-\frac{1}{2},0}
&=&\frac{2 \sqrt{3} \alpha_{\lambda_2}^{5/2}q
\alpha_3^{5/2} (\alpha_{\lambda_1} \alpha_{\rho_2}\alpha_{\rho_1})^{3/2}}{\pi^{9/4}
\left(\alpha_{\lambda_1}^2+\alpha_{\lambda_2}^2\right)^{5/2}
\left(\alpha_{\rho_1}^2+\alpha_{\rho_2}^2\right)^{3/2}} \frac{m_2}{2m_2+m_3'}
\left(\frac{1}{m_4}\mp\frac{1}{m_5}\right)\,,
\end{eqnarray}
where $(\alpha_{\rho_1},\alpha_{\lambda_1})$, $(\alpha_{\rho_2},\alpha_{\lambda_2})$,
and $\alpha_3$ are the oscillator parameters defined in Eq.~(\ref{SHO}),
corresponding to the initial-state baryon,
the final-state baryon, and the final-state meson, respectively.
Acting on the color and flavor wave functions, the operators
$\hat O_f$ and $\hat O_c$ yield
$\langle \zeta_{K^{*0}_0}|\delta_{c_4 c_5}|0\rangle
\langle\zeta_{N^0_{1535}}|\delta_{c_3' c_3}|\zeta_{\Lambda_b^0} \rangle=\sqrt{3}$
and $\langle \varphi_{K^{*0}_0}|b_5^\dagger(s)b_4^\dagger(\bar{d})|0\rangle$
$\langle \varphi_{N^0_{1535}} |b_3^\dagger(d)b_3(b) |\varphi_{\Lambda_b^0} \rangle=1$.
With these non-vanishing amplitudes incorporated into the total amplitude,
the corresponding branching fraction is evaluated through~\cite{Wang:2025pcs,Wang:2026dkd,
Wang:2022zja,Wang:2024ozz}
\begin{eqnarray}\label{Gamma}
&&
{\cal B}(\Lambda_b^0\to {\bf B}^* M)=8\pi^2\frac{|\textbf{q}|\tau_1 E_2 E_3}{M_{1}}
\sum_{J_1^z,J_2^z}|\mathcal{M}^{J_1^z,J_2^z,J_3^z}_{J_1,J_2,J_3}(\Lambda_b^0\to {\bf B}^* M)|^2\,,
\end{eqnarray}
where $\tau_1$ is the lifetime of the initial baryon, and the amplitudes
are summed over the spin components of the initial and final states.

The wave functions of $1P$-wave baryons can be found
in Refs.~\cite{Zhong:2024mnt,Koniuk:1979vy,Bijker:2015gyk}, while
those of the mesons ($K^-$, $\bar K_J^0$ and $M_J^0$)
can be obtained from, or constructed following,
Refs.~\cite{Wang:2025pcs,Afonin:2007aa,Pang:2017dlw,
Barnes:2002mu,Ebert:2009ub}.
Accordingly, the branching fractions of the remaining decay channels
$\Lambda_b^0\to N^*M,\Lambda^* M$ can also be evaluated within the CQM framework.
Using the approximate relations
${\cal B}(\Lambda_b^0 \to K^- p \pi^+\pi^-)$ $\simeq
{\cal B}(\Lambda_b^0 \to N^{*+}K^-){\cal B}(N^{*+}\to p\pi^+\pi^-)$,
${\cal B}(\Lambda_b^0 \to N^{*0} \bar K_J^{0})
{\cal B}(N^{*0}\to p\pi^-){\cal B}(\bar K_J^{0}\to K^-\pi^+)$, and
${\cal B}(\Lambda_b^0 \to \Lambda^* M_J^0)
{\cal B}(\Lambda^*\to p K^-){\cal B}(M_J^0\to\pi^+\pi^-)$,
we obtain the corresponding resonant branching fractions.
The direct $CP$ asymmetry is then defined as~\cite{LHCb:2025ray}
\begin{eqnarray}\label{ACP}
{\cal A}_{CP}(\Lambda_b^0 \to K^- p \pi^+\pi^-)\equiv
\frac{{\cal B}(\Lambda_b^0 \to K^- p \pi^+\pi^-)-{\cal B}(\bar\Lambda_b^0 \to K^+ \bar p \pi^-\pi^+)}
{{\cal B}(\Lambda_b^0 \to K^- p \pi^+\pi^-)+{\cal B}(\bar\Lambda_b^0 \to K^+ \bar p \pi^-\pi^+)}\,,
\end{eqnarray}
where $\bar\Lambda_b^0 \to K^+ \bar p \pi^-\pi^+$ denotes the corresponding antiparticle decay.

\section{Numerical analysis}
%
\begin{table}[b]
\caption{Branching fractions of the $N^*$ and $\Lambda^*$ resonances, and of the
$\Lambda_b^0\to N^* M$ and $\Lambda_b^0\to\Lambda^* M$ decays.
The first uncertainty arises from $a_i$, obtained by varying
$N_c^{\rm eff}$ from 2 to $\infty$ 
(and $a_2$ for the $\Lambda_b^0\to\Lambda^*(\rho^0,\omega)$ channels), 
together with the Wolfenstein parameters.
The second uncertainty arises from the baryon and meson oscillator parameters.}\label{tab1}
{\tiny
\begin{tabular}
{lcccc}
\hline
&$N_{1535}$&$N_{1520}$&$N_{1650}$&$N_{1700}$\\
\hline
$10^2{\cal B}(N^{*+}\to p\pi^+\pi^-)$
&$8.6\pm 5.2$
&$24.5\pm 3.0$
&$19.0\pm 7.0$
&$$$56.7\pm 6.0$\\
$10^6{\cal B}(\Lambda_b^0\to N^{*+} K^-)$
&$15.0^{+4.0+4.7}_{-1.8-4.1}$
&$25.9^{+6.9+7.7}_{-3.2-7.0}$
&$3.7^{+1.0+1.1}_{-0.5-1.0}$
&$4.0^{+1.1+1.2}_{-0.5-1.0}$\\
\hline
$10^2{\cal B}(N^{*0}\to p\pi^-)$
&$28.0\pm 6.7$
&$40.0\pm 3.3$
&$40.0\pm 6.7$
&$8.0\pm 3.3$\\
$10^6{\cal B}(\Lambda_b^0\to N^{*0} \bar K^{*0})$
&$6.8^{+2.6+2.0}_{-1.2-1.8}$
&$12.4^{+4.8+3.6}_{-2.1-3.2}$
&$1.7^{+0.7+0.5}_{-0.3-0.4}$
&$1.9^{+0.7+0.5}_{-0.3-0.5}$\\
$10^6{\cal B}(\Lambda_b^0\to N^{*0} \bar K^{*0}_0)$
&$15.4^{+2.8+5.7}_{-1.3-4.8}$
&$32.5^{+6.0+12.1}_{-2.8-10.2}$
&$3.9^{+0.7+1.4}_{-0.3-1.2}$
&$5.2^{+1.0+1.9}_{-0.5-1.6}$\\
\hline\hline
&$\Lambda_{1670}$&$\Lambda_{1690}$&$\Lambda_{1405}$&$\Lambda_{1520}$\\
\hline
$10^2{\cal B}(\Lambda^*\to p K^-)$
&$12.5\pm 2.5$
&$12.5\pm 2.5$
&$0$
&$22.5\pm 0.5$\\
$10^7{\cal B}(\Lambda_b^0\to \Lambda^*\rho^0)$
&$4.2^{+0.2+1.1}_{-0.2-1.0}$
&$7.2^{+0.4+1.8}_{-0.4-1.7}$
&$1.7^{+0.1+0.5}_{-0.1-0.4}$
&$3.2^{+0.2+0.8}_{-0.2-0.8}$\\
$10^6{\cal B}(\Lambda_b^0\to \Lambda^*\omega)$
&$0.9^{+0.5+0.2}_{-0.3-0.2}$
&$1.5^{+0.8+0.4}_{-0.5-0.3}$
&$0.4^{+0.2+0.1}_{-0.1-0.1}$
&$0.7^{+0.4+0.2}_{-0.2-0.2}$\\
$10^6{\cal B}(\Lambda_b^0\to \Lambda^* f_0)$
&$10.9^{+0.2+3.9}_{-0.2-3.3}$
&$22.1^{+0.4+7.9}_{-0.5-6.7}$
&$4.2^{+0.1+1.5}_{-0.1-1.3}$
&$9.4^{+0.2+3.4}_{-0.2-2.9}$\\
\hline
\end{tabular}
}
\end{table}
%
In the numerical analysis, the CKM matrix elements
in the Wolfenstein parameterization are written as
$(V_{ub},V_{tb})=(A\lambda^3(\rho-i\eta),\,1)$,
$(V_{ud},V_{td})=(1-\lambda^2/2,\,A\lambda^3)$, and
$(V_{us},V_{ts})=(\lambda,\,-A\lambda^2)$,
with $(\lambda,A)=(0.225,0.826)$
and $(\rho,\eta)=(0.163\pm 0.010,0.357\pm 0.010)$ taken from Ref.~\cite{pdg}.
For the oscillator parameters of initial-state $\Lambda_b^0$ baryon,
we adopt $\alpha_{\rho_1}=(400\pm 14)$~MeV from Ref.~\cite{Wang:2025pcs},
together with $\alpha_{\lambda_1}
= [3m_b/(m_u+m_d+m_b)]^{1/4}\alpha_{\rho_1}$~\cite{Wang:2017kfr,Yao:2018jmc}.
For the final-state (excited) baryon,
we use $\alpha_{\rho_2}=\alpha_{\lambda_2}=(606\pm 13)$~MeV,
extracted from the $\Lambda_b^0\to p\pi^-$ and $p K^-$ data.
From the same extraction, the final-state pion gives
$\alpha_3=\alpha_{\pi}=(750\pm 42)$~MeV, while the oscillator parameters
of the strange mesons are defined through
$\alpha_{M_{q\bar q}}=[2m_q m_{\bar q}/(m_q+m_{\bar q})m_u]^{1/2}
\alpha_\pi$~\cite{Zhong:2008kd}.
These extracted values are consistent with the results in Ref.~\cite{Kokoski:1985is}.
The constituent quark masses are chosen as $(m_u,m_d,m_s)$
=$(450,450,550)$~MeV, and $m_b=4.8$~GeV, following Refs.~\cite{Wang:2025pcs,Ni:2023lvx}.
The effective Wilson coefficients $c_i^{\rm eff}$ for the
$b\to sq\bar q'$ ($\bar b\to\bar s\bar q q'$) transition are given by~\cite{Ali:1998eb,Hsiao:2017tif}
\begin{eqnarray}\label{WCeff}
&&(c^{\rm eff}_1,c^{\rm eff}_2)=(1.168,-0.365)\,,\,
10^4 c^{\rm eff}_3=241.9\pm 3.2\eta + 1.4 \rho + i(31.3\mp 1.4\eta + 3.2\rho),\,\nonumber\\
&&10^4 c^{\rm eff}_4=-508.7 \mp 9.6\eta - 4.2\rho+ i(-93.9 \pm 4.2\eta - 9.6\rho) ,\,\nonumber\\
&&10^4 c^{\rm eff}_5=149.4\pm 3.2\eta + 1.4\rho + i(31.3\mp 1.4\eta + 3.2\rho),\,\nonumber\\
&&10^4 c^{\rm eff}_6=-645.5 \mp 9.6\eta- 4.2\rho + i(-93.9\pm 4.2\eta - 9.6\rho) ,\,\nonumber\\
&&10^4 c^{\rm eff}_9=-112.2 \mp 0.1\eta- 0.1\rho + i(-2.2\pm 0.1\eta - 0.1\rho) ,\,
10^4 c^{\rm eff}_{10}=37.5\,.
\end{eqnarray}
In Eq.~(\ref{amp1}), since $a_2$ is particularly sensitive to nonfactorizable QCD loop corrections,
the choice $N_c^{\rm eff}=3$ does not fully incorporate these effects.
Phenomenologically, the value $a_2\sim0.2$ is commonly applied in ${\bf B}_b$ decays,
corresponding to $N_c^{\rm eff}\simeq2$~\cite{Hsiao:2015txa,Hsiao:2015cda}.
In particular, we use $a_2=0.18\pm 0.05$ from Ref.~\cite{Hsiao:2017tif}.

The branching fractions of $\Lambda_b^0 \to N^{*+} K^-$, $N^{*0} \bar K_J^{0}$,
and $\Lambda^* M_J^0$ are calculated and listed in Table~\ref{tab1}.
The branching fractions of the $N^*$ ($\Lambda^*$) decays
are taken from the PDG~\cite{pdg} and summarized in Table~\ref{tab1}, together with
${\cal B}(\bar K^{*0},\bar K_0^{*0}\to K^-\pi^+)=(66.6,62.0\pm 6.7)\%$ and
${\cal B}(\rho^0,\omega,f_0\to\pi^+\pi^-)=(100,1.5\pm0.1,35\pm8)\%$~\cite{pdg}.
The $CP$ asymmetries of the resonant subprocesses are evaluated, and
the branching fraction and the corresponding $CP$ asymmetry
of the resonant four-body decay $\Lambda_b^0\to pK^-\pi^+\pi^-$ are obtained,
presented in Table~\ref{tab2}.

\section{Discussions and Conclusion}
%
\begin{table}[b]
\caption{Branching fractions and $CP$ asymmetries of resonant
$\Lambda_b^0\to pK^-\pi^+\pi^-$ channels.
Here, $N^*_{\rm sum}$ ($\Lambda^*_{\rm sum}$) denotes the sum over 
the four excited nucleon (hyperon) states in Table~\ref{tab1}, 
while $\bar K^0_{\rm sum}$ and $M^0_{\rm sum}$ denote 
the sums over $(\bar K^{*0},\bar K_0^{*0})$ and $(\rho^0,\omega,f_0)$, respectively.
The first two uncertainties follow Table~\ref{tab1}, 
and the third arises from the branching fractions of the intermediate resonances.}\label{tab2}
{\tiny
\begin{tabular}
{lcc}
\hline
resonant decay channel&${\cal B}\times 10^{6}$&${\cal A}_{CP}\times 10^2$\\
\hline
{\bf (measurement~\cite{LHCb:2025ray})}&&\\
$\Lambda_b^0\to K^- R(p\pi^+\pi^-)$
&------
&$5.4\pm 0.9\pm 0.1$\\
$\Lambda_b^0\to R(p\pi^-)R(K^-\pi^+)$
&------
&$2.7\pm 0.8\pm 0.1$\\
$\Lambda_b^0\to R(pK^-)R(\pi^+\pi^-)$
&------
&$5.3\pm 1.3\pm 0.2$\\
$\Lambda_b^0\to pK^-\pi^+\pi^-$
&------
&$2.45\pm0.46\pm0.10$\\
\hline\hline
{\bf (our work)}&&\\
$\Lambda_b^0\to K^-(N^{*+}_{\rm sum}\to)p\pi^+\pi^-$
&$10.6^{+1.8+2.1}_{-0.8-1.8}\pm1.2$
&$7.40\pm 0.15\pm 0.17\pm 0.20$\\
\hline
$\Lambda_b^0\to (N^{*0}_{\rm sum}\to)p\pi^-(\bar K^{*0}\to)K^-\pi^+$
&$5.1^{+1.4+1.0}_{-0.6-0.9}\pm0.4$
&$1.22\pm0.13\pm 0.01\pm 0.07$\\
$\Lambda_b^0\to (N^{*0}_{\rm sum}\to)p\pi^-(\bar K^{*0}_0\to)K^-\pi^+$
&$11.9^{+1.6+3.2}_{-0.7-2.7}\pm 1.3$
&$0.95\pm 0.14\pm 0.19\pm 0.14$\\
$\Lambda_b^0\to (N^{*0}_{\rm sum}\to)p\pi^-(\bar K_{\rm sum}^0\to)K^-\pi^+$
&$17.1^{+2.1+3.4}_{-1.0-2.9}\pm 1.4$
&$1.03\pm 0.16\pm 0.14\pm 0.19$\\
\hline
$\Lambda_b^0\to (\Lambda^*_{\rm sum}\to)pK^-(\rho^0\to)\pi^+\pi^-$
&$0.21\pm 0.01\pm 0.03\pm 0.02$
&$1.37\pm 0.00\pm 0.08^{+0.14}_{-0.06}$\\
$\Lambda_b^0\to (\Lambda^*_{\rm sum}\to)pK^-(\omega\to)\pi^+\pi^-$
&$(0.7^{+0.2}_{-0.1}\pm 0.1 \pm 0.1)\times 10^{-2}$
&$6.29^{+0.31+0.03+0.06}_{-0.50-0.03-0.05}$\\
$\Lambda_b^0\to (\Lambda^*_{\rm sum}\to)pK^-(f_0\to)\pi^+\pi^-$
&$2.2\pm 0.0^{+0.5}_{-0.4}\pm 0.4$
&$0.94^{+0.01+0.99+0.05}_{-0.02-1.21-0.10}$\\
$\Lambda_b^0\to (\Lambda^*_{\rm sum}\to)pK^-(M^0_{\rm sum}\to)\pi^+\pi^-$
&$2.4\pm 0.0^{+0.5}_{-0.4}\pm0.4$
&$0.99\pm 0.08^{+0.97}_{-1.24}\pm 0.11$\\
\hline
$\Lambda_b^0\to pK^-\pi^+\pi^-$
&$30.0^{+2.8+4.0}_{-1.3-3.4}\pm1.8$
&$3.18\pm 0.11\pm0.13 \pm 0.11$\\
\hline
\end{tabular}
}
\end{table}
%
Experimentally, the subprocesses
$\Lambda_b^0\to R(p\pi^+\pi^-)K^-$,
$\Lambda_b^0\to R(p\pi^-)R(K^-\pi^+)$, and
$\Lambda_b^0\to R(pK^-)R(\pi^+\pi^-)$
have been identified in the four-body decay
$\Lambda_b^0\to pK^-\pi^+\pi^-$.
We interpret the first subprocess as
$\Lambda_b^0\to K^-(N^{+}_{\rm sum}\to)p\pi^+\pi^-$,
where $N^{+}_{\rm sum}$ denotes the combined contribution of the participating $N^*$ states.
For the first time, the relevant resonances are identified as
$N^+_{1535}$, $N^+_{1520}$, $N^+_{1650}$, and $N^+_{1700}$.
We predict ${\cal B}(\Lambda_b^0\to K^-(N^{*+}_{\rm sum}\to)p\pi^+\pi^-)
=(10.6^{+1.8+2.1}_{-0.8-1.8}\pm1.2)\times10^{-6}$, dominated by the $N^+_{1520}$ contribution.
Within the CQM framework, the spin structure of the $\Lambda_b^0$ baryon 
has no overlap with that of $N_{1675}$. 
As a result, the $\Lambda_b^0\to N_{1675}$ transition vanishes, 
yielding ${\cal B}(\Lambda_b^0\to N_{1675}M)=0$.
Since the branching fractions ${\cal B}(\Lambda_b^0\to N^+_{1535,1520}K^-)$
are at the $10^{-5}$ level, significantly larger than
${\cal B}(\Lambda_b^0\to p K^-)=(5.5\pm1.0)\times10^{-6}$ for the ground-state nucleon~\cite{pdg},
this enhancement provides a useful test of the applicability of the CQM framework.
Remarkably, the resulting
${\cal A}_{CP}^+\equiv{\cal A}_{CP}(\Lambda_b^0\to K^-(N^{*+}_{\rm sum}\to)p\pi^+\pi^-)
=(7.40\pm 0.15\pm 0.17\pm 0.20)\%$ is in good agreement with
the experimental value $(5.4\pm 0.9\pm 0.1)\%$.

The subprocess $\Lambda_b^0\to R(p\pi^-)R(K^-\pi^+)$ is interpreted as
$\Lambda_b^0\to (N^{*0}_{\rm sum}\bar K^{*0},N^{*0}_{\rm sum}\bar K_0^{*0})$,
followed by $(\bar K^{*0},\bar K_0^{*0})\to K^-\pi^+$, where
$N^{*0}_{\rm sum}$ denotes the combined contribution of
$N^0_{1535}$, $N^0_{1520}$, $N^0_{1650}$, and $N^0_{1700}$. We obtain
${\cal B}(\Lambda_b^0\to (N^{*0}_{\rm sum}\to)p\pi^-[(\bar K^{*0},\bar K_0^{*0})\to]K^-\pi^+)
=(5.1^{+1.4+1.0}_{-0.6-0.9}\pm0.4,11.9^{+1.6+3.2}_{-0.7-2.7}\pm1.3)\times10^{-6}$.
The corresponding $CP$ asymmetries, ${\cal A}_{CP}(\bar K^{*0},\bar K_0^{*0})$
$=(1.22\pm0.13\pm0.01\pm0.07,0.95\pm0.14\pm0.19\pm0.14)\%$, 
are several times smaller than ${\cal A}_{CP}^{+}\simeq 6\%$. 
As indicated by Eq.~(\ref{amp1}), the decays
$\Lambda_b^0\to N^{*0}\bar K^{*0}$ and $\Lambda_b^0\to N^{*0}\bar K_0^{*0}$
are dominated by penguin amplitudes and receive no tree-level contributions 
carrying the weak phase associated with $V_{ub}$.
The resulting absence of tree--penguin interference naturally accounts for the suppression of
${\cal A}_{CP}(\bar K^{*0},\bar K_0^{*0})$ relative to ${\cal A}_{CP}^{+}$.
Summing over $(\bar K^{*0},\bar K_0^{*0})\equiv \bar K^0_{\rm sum}$ yields
${\cal A}_{CP}(\Lambda_b^0\to (N^{*0}_{\rm sum}\to)p\pi^-
(\bar K_{\rm sum}^0\to)K^-\pi^+)=(1.03\pm0.16\pm0.14\pm0.19)\%$,
consistent with the experimental value $(2.7\pm 0.8\pm 0.1)\%$.

The subprocess $\Lambda_b^0\to R(pK^-)R(\pi^+\pi^-)$ is interpreted as
$\Lambda_b^0\to (\Lambda^*_{\rm sum}\to)pK^-(M_J^0\to)\pi^+\pi^-$,
where $\Lambda^*_{\rm sum}$ denotes the combined contribution of
$\Lambda_{1670}$, $\Lambda_{1690}$, and $\Lambda_{1520}$, and
$M_J^0=(\rho^0,\omega,f_0)$. The $\Lambda_{1405}$ resonance is not involved, 
as its mass lies below the $pK^-$ threshold, rendering
$\Lambda_{1405}\to pK^-$ kinematically forbidden~\cite{pdg}.
Since $\rho^0=(u\bar u-d\bar d)/\sqrt 2$, one has
$\langle\rho^0|(\bar uu+\bar dd)_{V-A}|0\rangle=0$, 
leading to the suppressed amplitude
$\hat{\cal M}(\Lambda_b^0\to\Lambda^*\rho^0)\propto\alpha_2^s+3\alpha_9^s$.
Defining ${\cal B}(M_J^0)\equiv
{\cal B}(\Lambda_b^0\to(\Lambda^*_{\rm sum}\to)pK^-(M_J^0\to)\pi^+\pi^-)$,
we obtain ${\cal B}(\rho^0)=(0.21\pm0.01\pm0.03\pm0.02)\times10^{-6}$.
Moreover, since $\alpha_9^s$ acquires only a negligible strong phase from
$(c_9^{\rm eff},c_{10}^{\rm eff})$ [see Eq.~(\ref{WCeff})], its interference with the weak phase in $\alpha_2^s$ is weak, resulting in the small asymmetry ${\cal A}_{CP}(\rho^0)=(1.37\pm0.00\pm0.08^{+0.14}_{-0.06})\%$.

Unlike the $\rho^0$ channel, the amplitude
${\cal M}(\Lambda_b^0\to\Lambda^*\omega)\propto
\alpha_2^s+(2\alpha_3^s+2\alpha_5^s+\alpha_9^s)$, with $\omega=(u\bar u+d\bar d)/\sqrt2$,
does not suffer from flavor suppression.
Nevertheless, the branching fraction remains small, ${\cal B}(\omega)\simeq
{\cal B}(\Lambda_b^0\to\Lambda^*_{\rm sum} \omega){\cal B}(\omega\to\pi^+\pi^-)
=(0.7^{+0.2}_{-0.1}\pm0.1\pm 0.1)\times 10^{-8}$, due to the suppressed decay $\omega\to\pi^+\pi^-$.
The sizable penguin-induced strong phases associated with $\alpha_3^s$ and $\alpha_5^s$, 
together with the weak phase in the tree amplitude, lead to ${\cal A}_{CP}(\omega)\simeq 6\%$.
The $f_0$ state with $J^{PC}=0^{++}$ receives contributions solely from the
$s\bar s$ scalar current associated with the $\alpha_6^s$ term.
Adopting the mixing scheme $|f_0\rangle=\cos\theta|s\bar s\rangle
+\sin\theta|u\bar u+d\bar d\rangle/\sqrt{2}$, 
with $\theta=(156.7\pm 0.7)^\circ$~\cite{Hsiao:2023qtk},
we obtain ${\cal B}(f_0)=(2.2\pm0.0^{+0.5}_{-0.4}\pm0.4)\times10^{-6}$.
The corresponding asymmetry, ${\cal A}_{CP}(f_0)\simeq 1\%$, is comparable to
${\cal A}_{CP}(\bar K^{*0},\bar K_0^{*0})$, reflecting the absence of tree-level amplitudes 
carrying the weak phase associated with $V_{ub}$.
Summing over all $M_J^0$ contributions yields
${\cal A}_{CP}(\Lambda_b^0\to(\Lambda^*_{\rm sum}\to)pK^-(M^0_{\rm sum}\to)\pi^+\pi^-)
=(0.99\pm0.08^{+0.97}_{-1.24}\pm0.11)\%$, which differs from the current measurement,
$(5.3\pm1.3\pm0.2)\%$, by $2.6\sigma$.

Combining the contributions from $N^*$, $\Lambda^*$, 
and the relevant meson resonances, we obtain
${\cal B}(\Lambda_b^0\to pK^-\pi^+\pi^-)
=(30.0^{+2.8+4.0}_{-1.3-3.4}\pm1.8)\times 10^{-6}$ and
${\cal A}_{CP}(\Lambda_b^0\to pK^-\pi^+\pi^-)
=(3.18\pm0.11\pm 0.13\pm 0.11)\%$,
consistent with the measured value
$(2.45\pm0.46\pm0.10)\%$~\cite{LHCb:2025ray}.
In contrast to the penguin-dominated modes studied here, 
the corresponding tree-dominated channels are also of considerable interest. 
Focusing on the dominant subprocess
$\Lambda_b^0\to\pi^-(N^{*+}_{\rm sum}\to)p\pi^+\pi^-$, we predict
${\cal B}(\Lambda_b^0\to p\pi^-\pi^+\pi^-)
=(8.3^{+1.3+1.7}_{-0.6-1.5}\pm0.9)\times10^{-6}$ and
${\cal A}_{CP}(\Lambda_b^0\to p\pi^-\pi^+\pi^-)
=(-5.90\pm 0.14^{+0.01}_{-0.03}\pm0.02)\%$,
which can be tested at LHCb.

In this work, we have investigated the resonant four-body decay
$\Lambda_b^0\to pK^-\pi^+\pi^-$ and the underlying two-body transitions
$\Lambda_b^0\to N^*M$ and $\Lambda_b^0\to\Lambda^*M$ within the CQM framework.
The dominant contributions have been identified as the $1P$-wave baryon states
$N^*=(N_{1535},N_{1520},N_{1650},N_{1700})$ and
$\Lambda^*=(\Lambda_{1670},\Lambda_{1690},\Lambda_{1520})$.
By combining the contributions from these excited baryons and the relevant meson resonances, 
we have obtained ${\cal B}(\Lambda_b^0\to pK^-\pi^+\pi^-)
=(30.0^{+2.8+4.0}_{-1.3-3.4}\pm1.8)\times10^{-6}$ and
${\cal A}_{CP}(\Lambda_b^0\to pK^-\pi^+\pi^-)=(3.18\pm0.11\pm0.13\pm0.11)\%$,
with the latter in remarkable agreement with experiment.
We have further predicted ${\cal B}(\Lambda_b^0\to p\pi^-\pi^+\pi^-)
=(8.3^{+1.3+1.7}_{-0.6-1.5}\pm0.9)\times10^{-6}$ and ${\cal A}_{CP}(\Lambda_b^0\to p\pi^-\pi^+\pi^-)
=(-5.90\pm0.14^{+0.01}_{-0.03}\pm0.02)\%$, which can be tested at LHCb.
We have thus established the first comprehensive framework 
for systematically identifying the excited $N^*$ and $\Lambda^*$ resonances 
relevant to multi-body beauty-baryon decays and for
quantitatively evaluating their impact on branching fractions and baryonic $CP$ asymmetries.

\section*{ACKNOWLEDGMENTS}
YKH was supported by the National Natural Science Foundation of China (NSFC)
under Grants No.~12575101 and No.~12175128.
KLW was supported by the NSFC under Grant No.~12205026,
and by the Support Project for Young Teachers' Research and Innovation Abilities
under Grant No.~2025Q037.
JW was supported by the Shanxi Provincial Graduate Education Innovation Program Project
under Grant No.~2025XS111.


\end{document}